\documentclass[11pt,a4paper]{article}

\usepackage[T1]{fontenc}
\usepackage[utf8]{inputenc}
\usepackage{lmodern}
\usepackage[margin=2.8cm]{geometry}
\usepackage{amsmath,amssymb,amsthm}
\usepackage{enumitem}
\usepackage{parskip}
\usepackage{microtype}
\usepackage{tikz}
\usetikzlibrary{calc}
\usepackage[hidelinks]{hyperref}

\newtheorem{theorem}{Theorem}
\newtheorem{lemma}[theorem]{Lemma}
\newtheorem{proposition}[theorem]{Proposition}
\newtheorem{flawedclaim}[theorem]{Flawed claim}
\theoremstyle{remark}

\theoremstyle{definition}
\newtheorem{definition}[theorem]{Definition}

\newcommand{\bbF}{\mathbb{F}}
\newcommand{\bbN}{\mathbb{N}}
\newcommand{\RS}{\mathrm{RS}}
\newcommand{\Fold}{\mathrm{Fold}}
\newcommand{\Cut}{\mathrm{Cut}}
\newcommand{\Cay}{\mathrm{Cay}}
\newcommand{\C}{\mathcal{C}}
\newcommand{\W}{W}
\newcommand{\dV}{\Delta_V}
\newcommand{\dH}{\Delta_H}
\newcommand{\U}{U}
\newcommand{\one}{\mathbf{1}}

\title{A counterexample to the soundness of the Flowering IOPP\\ for codes on graphs}
\author{Hugo Delavenne\\[2pt]
  \small LIX, \'Ecole polytechnique, Institut Polytechnique de Paris\\\small Inria}
\date{}

\begin{document}
\maketitle

\begin{abstract}
  The Flowering protocol of~\cite{DMR25} and its generalization in~\cite{DL25} are Interactive Oracle Proofs of Proximity (IOPPs) for codes on graphs with Reed-Solomon local codes.
  Similarly to the FRI for Reed-Solomon codes, the graph is recursively cut into isomorphic subgraphs and the word folded along the cuts.
  We show here that the soundness claims of~\cite{DMR25,DL25} are false.
  The commit soundness proofs identify the vertex distance of a word to the code with the fraction of its unsatisfied local constraints, while only one inequality holds in the wrong direction.
  We furthermore give an explicit counterexample, which shows that this approach cannot create an efficient IOPP.
\end{abstract}

\section{Introduction}

IOPPs like the FRI protocol~\cite{BBHR18} successively reduce a claim of the form $\Delta(f,C)\leq\delta$ for $C\subseteq\bbF^n$, $f\in\bbF^n$ and $\delta\in[0,1]$ to a claim $\Delta(f',C')\leq\delta-\eta$ with $\eta>0$, $C'\subseteq\bbF^{n/2}$ and $f'\in\bbF^{n/2}$ is obtained from $f$ and public coin randomness from the Verifier, and is called the \emph{folding} of $f$ by the public coin randomness.
The soundness analysis of these protocols then relies on the fact that with high probability on the public coin of the Verifier, the folding of a vector $f$ $\delta$-far from the code $C$ is $(\delta-\eta)$-far from~$C'$.

As we recall in Section~\ref{sec:setting}, in~\cite{DMR25} the authors propose an analogous IOPP called Flowering, for codes on graphs~\cite{Tan81,SS96} whose coordinates are the edges of a regular graph and whose local views must lie in a Reed-Solomon code.
The vertices of the graph are cut into two subsets of vertices, and the subvectors on the induced subgraphs are combined linearly.
This construction is extended in~\cite{DL25} to collections of non-disjoint cuts.
Both papers claim a soundness error smaller than that of FRI, valid up to the covering radius.
However these soundness claims are false.

In Section~\ref{sec:counterexamples}, we locate the flaw in the proofs of commit soundness \cite[Proposition~3]{DMR25} and \cite[Proposition~4]{DL25}, we show that folding along a flowering cut does not preserve the Hamming distance, and we give a word on which a cheating Prover contradicts the soundness theorems of both papers.

\section{Definitions and claim}\label{sec:setting}

We follow the definitions and notations of~\cite{DMR25}.

\begin{definition}[$n$-RIM]
  For $n\in\bbN$, an \emph{$n$-regular indexed multigraph} ($n$-RIM) $\Gamma=(V,E)$ is given by a map $E\colon V\times[n]\to V$ such that for all $v\in V$ and $\ell\in[n]$, $E(E(v,\ell),\ell)=v$.
\end{definition}

\noindent
We say that $(v,\ell)$ is a \emph{petal} if $E(v,\ell)=v$.
A \emph{word} on $\Gamma$ is a map $f\colon V\times[n]\to\bbF$ such that for all $v\in V$ and $\ell\in[n]$, $f(v,\ell)=f(E(v,\ell),\ell)$.
Denote $\W(\Gamma,\bbF)$ the set of words on $\Gamma$.
For~$f\in\W(\Gamma,\bbF)$ and $v\in V$, denote $f(v,\cdot)=(f(v,1),\dots,f(v,n))\in\bbF^n$ the \emph{local view} of~$f$ at~$v$.

\begin{definition}[Code {$\C[\Gamma,k]$}]
  Let $k\le n$ and let $\RS[n,k]$ be a Reed-Solomon code of length $n$ and dimension $k$.
  The code on $\Gamma$ is defined by
  \[
    \C[\Gamma,k]:=\{f\in\W(\Gamma,\bbF) \mid \forall v\in V, f(v,\cdot)\in\RS[n,k]\}.
  \]
\end{definition}

\noindent
Denote $\dH$ the relative Hamming distance, i.e. for $f,g\in\W(\Gamma,\bbF)$, $\dH(f,g)$ is the fraction of edges on which $f$ and $g$ differ.
For $f\in \W(\Gamma,\bbF)$ and $S\subseteq\W(\Gamma,\bbF)$, $\dH(f,S):=\nobreak\min_{g\in S}\dH(f,g)$.
For $x\in\bbF^n$, denote $|x|$ its Hamming weight.

\begin{definition}[Cut-graph]
  For $V'\subseteq V$, the \emph{cut-graph} $\Cut[\Gamma,V']=(V',E')$ is defined by $E'(v,\ell)=E(v,\ell)$ if $E(v,\ell)\in V'$ and $E'(v,\ell)=v$ otherwise.
  The edges leaving $V'$ become petals.
\end{definition}

\begin{definition}[Flowering cut]
  A \emph{flowering cut} is a (not necessarily disjoint) decomposition $V=V'\cup V''$ together with a bijection $\varphi\colon V'\to V''$ that is an index-preserving isomorphism from $\Cut[\Gamma,V']$ to $\Cut[\Gamma,V'']$, and such that for all $v\in V'$, $v\in V''$ if and only if $\varphi(v)\in V'$.
\end{definition}

The last condition is the invariance of the multiplicity \cite[Equation~(2)]{DL25}.

\begin{definition}[Folding]
  For $\alpha\in\bbF$, the \emph{folding} of $f\in\W(\Gamma,\bbF)$ is the word $\Fold[f,\alpha]\in\W(\Cut[\Gamma,V'],\bbF)$ given for $v\in V'$ and $\ell\in[n]$ by
  \[
    \Fold[f,\alpha](v,\ell)=f(v,\ell)+\alpha f(\varphi(v),\ell)\text.
  \]
\end{definition}

\begin{definition}[Blossoming sequence]
  A \emph{blossoming sequence} is a sequence of $n$-RIMs $\Gamma_0,\dots,\Gamma_r$ such that $\Gamma_r$ has a single vertex and each $\Gamma_i$ is the cut-graph $\Cut[\Gamma_{i-1},V_i]$ of a flowering cut $(V_i,\varphi_i)$ of $\Gamma_{i-1}$.
\end{definition}

\noindent
For a blossoming sequence, denote $\C_i:=\C[\Gamma_i,k]$ and $V_i'':=\varphi_i(V_i)$; when applied to a word on $\Gamma_{i-1}$, $\Fold$ denotes the folding along $(V_i,\varphi_i)$.

We present the Flowering protocol of~\cite[Definition~10]{DL25} for cuts of order~$2$ and with parameter $\mu=1$, that is, the Verifier reads whole local views.
For disjoint cuts, it is the protocol of~\cite[Protocol~1]{DMR25} with parameter $t=n$.

\begin{definition}[Flowering protocol~{\cite[Definition~10]{DL25}}]
  Let $m\geq 1$ and $f_0\in\W(\Gamma_0,\bbF)$.
  The \emph{Flowering protocol} is the following interactive protocol between a Prover and a Verifier where the Prover has access to $f_0$ and the Verifier has oracle access to $f_0$.
  \begin{itemize}[leftmargin=*]
  \item \emph{Commit phase.} For $i=1,\dots,r$, the Verifier sends a uniformly random $\alpha_{i-1}\in\bbF$, and the Prover answers with oracle access to some $f_i\in\W(\Gamma_i,\bbF)$.
  \item \emph{Query phase}, repeated independently $m$ times.
  The Verifier picks a uniformly random~$v_0\in\nobreak V_0$.
  For~$i\in[r]$, it sets $v_i:=v_{i-1}$ if $v_{i-1}\in V_i\setminus V_i''$, $v_i:=\varphi_i^{-1}(v_{i-1})$ if $v_{i-1}\in V_i''\setminus V_i$, and otherwise picks $v_i\in\{v_{i-1},\varphi_i^{-1}(v_{i-1})\}$ by choosing one of the two cuts uniformly at random.
    It checks that $\Fold[f_{i-1},\alpha_{i-1}](v_i,\cdot)=f_i(v_i,\cdot)$ for all $i\in[r]$, and that~$f_r(v_r,\cdot)\in\RS[n,k]$.
  \end{itemize}
\end{definition}

The main result of~\cite{DMR25,DL25} is the soundness of the Flowering protocol \cite[Theorem~2]{DL25}, which generalizes \cite[Theorem~2]{DMR25} to collections of non-disjoint cuts.
A flowering cut as above is a flowering cut collection of order~$2$ in the sense of \cite[Definition~7]{DL25}.
With $r$ rounds and $m$ repetitions, \cite[Theorem~2]{DL25} then reads as follows.

\begin{flawedclaim}[{\cite[Theorem~2]{DL25}}]\label{claim}
  Let $k<n$, let $(\Gamma_0,\dots,\Gamma_r)$ be a blossoming sequence and let $f_0\in\W(\Gamma_0,\bbF)$.
  For every Prover, the Verifier of the Flowering protocol accepts $f_0$ with probability at most
  \[
    \min_{\varepsilon>0}\Bigl(\frac{r}{\varepsilon|\bbF|}+\bigl(1-\dH(f_0,\C_0)+r\varepsilon\bigr)^m\Bigr).
  \]
\end{flawedclaim}

\noindent
We give a counterexample to Flawed claim~\ref{claim} in Section~\ref{sec:protocol}.

\section{Counterexamples}\label{sec:counterexamples}

For $f\in\W(\Gamma,\bbF)$, let $\U(f):=\{v\in V: f(v,\cdot)\notin\RS[n,k]\}$ be its set of \emph{unsatisfied vertices}.
Both papers prove Flawed claim~\ref{claim} through a commit-soundness statement~\cite[Proposition~3]{DMR25}, \cite[Proposition~4]{DL25} in the vertex distance $\dV(f,g):=|\{v\in V: f(v,\cdot)\neq g(v,\cdot)\}|/|V|$.
Its proof sets $\delta:=\dV(f,\C)$ and $T:=\U(f)$, and uses $|T|=\delta|V|$ (just before~\cite[Equation~(3)]{DMR25}, just after~\cite[Equation~(6)]{DL25}).
Only $|T|\le\delta|V|$ holds: a codeword differs from $f$ at every unsatisfied vertex, and possibly at many others (Section~\ref{sec:protocol}).

\subsection{Folding does not preserve the Hamming distance}\label{sec:counterexample}

We first show that folding may drastically decrease the Hamming distance, which excludes the possibility to have a proof similar to that of FRI.

\begin{figure}[ht]
  \centering
  \begin{tikzpicture}[
    vertex/.style={circle,fill=black,inner sep=1.6pt},
    val/.style={font=\scriptsize,inner sep=1pt},
    petal/.style={semithick},
    bad/.style={very thick,red}]
    \coordinate (a) at (0,0);
    \coordinate (b) at (2.4,0);
    \foreach \t in {120,160,200,240} {
      \draw[petal] (a) .. controls +(\t-20:1) and +(\t+20:1) .. (a);
      \node[val] at ($(a)+(\t:1.0)$) {$0$};
    }
    \foreach \t in {60,20,-20,-60} {
      \draw[petal] (b) .. controls +(\t-20:1) and +(\t+20:1) .. (b);
      \node[val] at ($(b)+(\t:1.0)$) {$1$};
    }
    \draw[petal] (a) -- node[val,above]{$0$} (b);
    \node[vertex,label={[font=\small]-60:$v_1$}] at (a) {};
    \node[vertex,label={[font=\small]240:$v_2$}] at (b) {};
    \node[font=\small] at (1.2,-1.6) {$\Gamma_0$ and $f$};
    \draw[->,thick] (4.3,0) -- node[above,font=\small]{$\Fold[\,\cdot\,,\alpha]$} (5.9,0);
    \coordinate (c) at (7.6,0);
    \foreach \t in {90,162,234,306} {
      \draw[petal] (c) .. controls +(\t-20:1) and +(\t+20:1) .. (c);
      \node[val] at ($(c)+(\t:1.0)$) {$\alpha$};
    }
    \draw[bad] (c) .. controls +(-2:1) and +(38:1) .. (c);
    \node[val,text=red] at ($(c)+(18:1.0)$) {$0$};
    \node[vertex,label={[font=\small,label distance=0.5cm]270:$v_1$}] at (c) {};
    \node[font=\small] at (7.6,-1.6) {$\Gamma_1$ and $\Fold[f,\alpha]$};
  \end{tikzpicture}
  \caption{Proposition~\ref{prop:hamming} for $n=5$.
  The common edge becomes the red petal, the only edge where $\Fold[f,\alpha]$ differs from the codeword $\alpha(1,\dots,1)$.}
  \label{fig:hamming}
\end{figure}
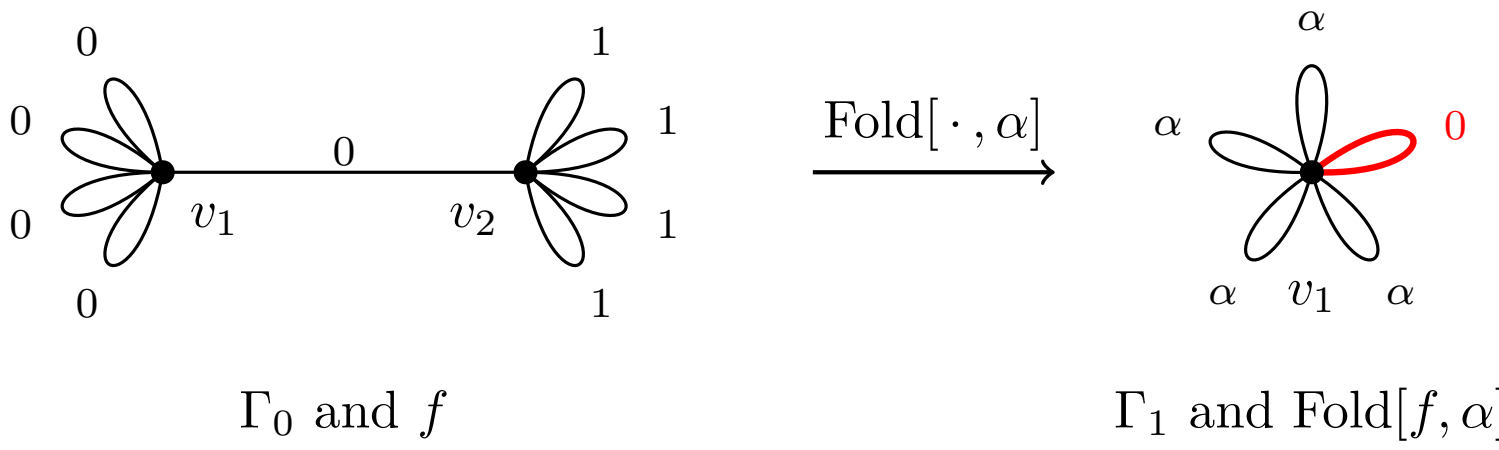

\begin{proposition}\label{prop:hamming}
  Let $n\ge3$, $|\bbF|>n$, $k=1$, and let $\Gamma_0$ have two vertices $v_1,v_2$ sharing the edge of index~$1$, each with $n-1$ petals.
  With $V':=\{v_1\}$ and $\varphi(v_1):=v_2$, the word $f$ given by $f(v_1,\cdot)=0$ and $f(v_2,\cdot)=(0,1,\dots,1)$ satisfies, for every $\alpha\in\bbF$,
  \[
    \dH(f,\C_0)=\frac{n-1}{2n-1}
    \qquad\text{and}\qquad
    \dH(\Fold[f,\alpha],\C_1)\le\frac1n
    \text.
  \]
\end{proposition}

\begin{proof}
  Since $\RS[n,1]$ is the set of constant vectors and $v_1,v_2$ share an edge, $\C_0$ is the set of constant words.
  The word $f$ equals $1$ on the $n-1$ petals of $v_2$ and $0$ on the $n$ other edges.
  For $x\in\bbF$, denote $f_x$ the constant word equal to $x$.
  Hence $f$ is at distance $n-1$ from $f_0$, $n$ from $f_1$, and $2n-1$ from $f_x$ for other values of $x$.
  Therefore $\dH(f,\C_0)=(n-1)/(2n-1)$.

  Moreover,
  \[
    \Fold[f,\alpha](v_1,\cdot)=f(v_1,\cdot)+\alpha f(v_2,\cdot)=\alpha(0,1,\dots,1),
  \]
  which differs from $\alpha(1,\dots,1)\in\C_1$ on a single edge (except for $\alpha=0$ in which case they are equal).
\end{proof}

\subsection{The claim is false}\label{sec:protocol}

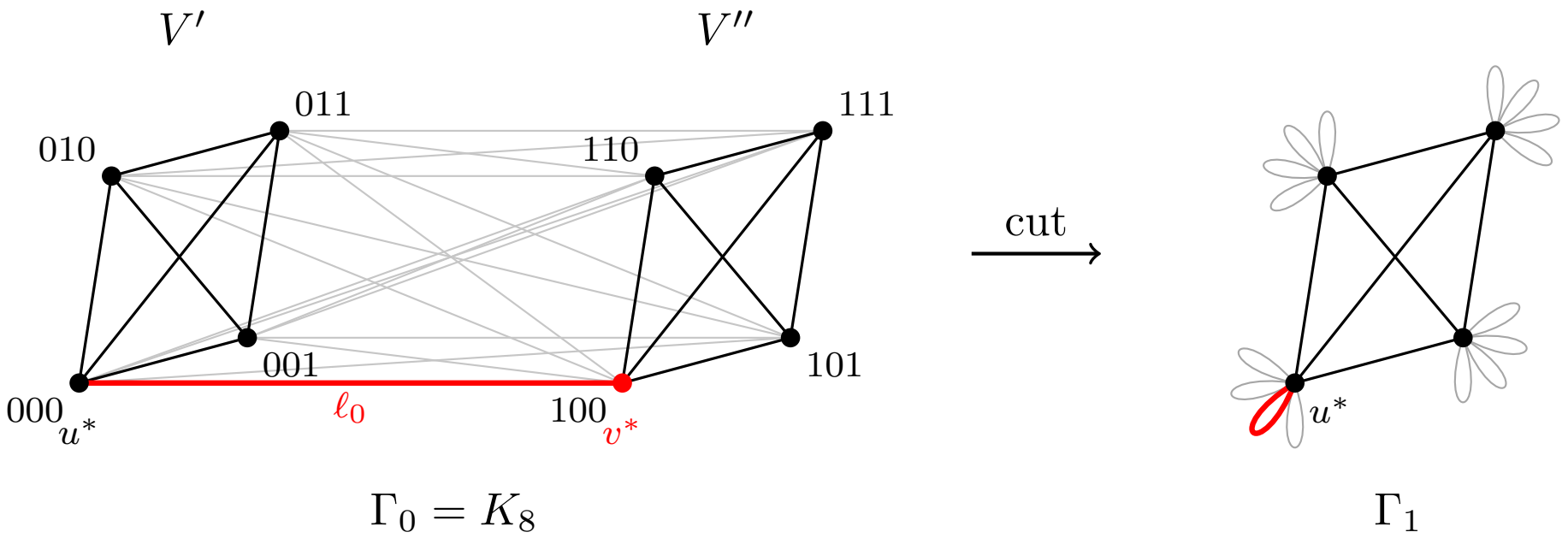
\begin{figure}[ht]
  \centering
  \begin{tikzpicture}[
    vertex/.style={circle,fill=black,inner sep=1.5pt},
    lab/.style={font=\scriptsize,inner sep=1.5pt},
    inner/.style={semithick},
    cross/.style={gray!45},
    petal/.style={gray!70},
    bad/.style={very thick,red}]
    \coordinate (A) at (0,0);     \coordinate (B) at (1.3,0.35);
    \coordinate (C) at (0.25,1.6); \coordinate (D) at (1.55,1.95);
    \coordinate (A') at (4.2,0);     \coordinate (B') at (5.5,0.35);
    \coordinate (C') at (4.45,1.6); \coordinate (D') at (5.75,1.95);
    \foreach \x in {A,B,C,D} \foreach \y in {A',B',C',D'} \draw[cross] (\x) -- (\y);
    \foreach \x/\y in {A/B,A/C,A/D,B/C,B/D,C/D} {
      \draw[inner] (\x) -- (\y);
      \draw[inner] (\x') -- (\y');
    }
    \draw[bad] (A) -- node[lab,below,text=red]{$\ell_0$} (A');
    \foreach \p/\l/\d in {A/000/below left,B/001/below right,C/010/above left,D/011/above right,
      B'/101/below right,C'/110/above left,D'/111/above right}
    \node[vertex,label={[lab]\d:$\l$}] at (\p) {};
    \node[vertex,fill=red,label={[lab]below left:$100$}] at (A') {};
    \node[lab,below=6pt] at (A) {$u^*$};
    \node[lab,below=6pt,text=red] at (A') {$v^*$};
    \node[font=\small] at (0.8,2.75) {$V'$};
    \node[font=\small] at (5.0,2.75) {$V''$};
    \node[font=\small] at (2.9,-1.0) {$\Gamma_0=K_8$};
    \draw[->,thick] (6.9,1) -- node[above,font=\small]{cut} (7.9,1);
    \begin{scope}[xshift=9.4cm]
      \coordinate (A) at (0,0);     \coordinate (B) at (1.3,0.35);
      \coordinate (C) at (0.25,1.6); \coordinate (D) at (1.55,1.95);
      \foreach \p/\angles in {A/{150,190,230,270},B/{-90,-50,-10,30},C/{90,130,170,210},D/{-30,10,50,90}}
      \foreach \t in \angles
      \draw[petal] (\p) .. controls +(\t-18:0.7) and +(\t+18:0.7) .. (\p);
      \draw[bad] (A) .. controls +(212:0.7) and +(248:0.7) .. (A);
      \foreach \x/\y in {A/B,A/C,A/D,B/C,B/D,C/D} \draw[inner] (\x) -- (\y);
      \foreach \p in {A,B,C,D} \node[vertex] at (\p) {};
      \node[lab,anchor=north west] at ($(A)+(0.05,-0.05)$) {$u^*$};
      \node[font=\small] at (0.8,-1.0) {$\Gamma_1$};
    \end{scope}
  \end{tikzpicture}
  \caption{
    First cut for $r=3$ and $s_{\ell_0}=e_1$.
    The gray edges cross the cut and become petals of~$\Gamma_1$.
    The only unsatisfied vertex of~$f$ is~$v^*$, so for $\alpha\neq0$ the only unsatisfied vertex of~$\Fold[f,\alpha]$ is~$u^*$.
    Its local view is corrected by changing the value on the red petal~$(u^*,\ell_0)$, which no other local view contains, and this yields a codeword of~$\C_1$ (Proposition~\ref{prop:main}).
  }
  \label{fig:k8}
\end{figure}

Fix $r\geq 3$, $n:=2^r-1$ and $k:=2^{r-1}$, so that $\RS[n,k]$ has minimum distance $2^{r-1}$.
Fix a field $\bbF$ of odd characteristic with $|\bbF|>n$.
Let $\Gamma_0:=\Cay(\bbF_2^r,S)$ with $S:=\bbF_2^r\setminus\{0\}=\{s_1,\dots,s_n\}$, that is $E(v,\ell):=v+s_\ell$.
Then $\Gamma_0$ is the complete graph on $2^r$ vertices.
Let $\Gamma_0,\dots,\Gamma_r$ be the blossoming sequence of \cite[Definition~10]{DMR25}: $V_i:=\{0\}^i\times\bbF_2^{r-i}$ and $\varphi_i(v):=v+e_i$, with $(e_1,...,e_r)$ the canonical basis.
Denote $V':=V_1$, $V'':=V_0\setminus V_1$ and $\varphi:=\varphi_1$.
For $u\in V'$ and $\ell\in[n]$, the vertex $u+s_\ell$ is in $V''$ if and only if the first coordinate of $s_\ell$ is $1$.
Hence, by denoting
\[
  L_\times:=\{\ell\in[n]\mid\text{the first coordinate of }s_\ell\text{ is }1\},
\]
the pair $(u,\ell)$ is a petal of $\Gamma_1$ for all $u\in V'$ and $\ell\in L_\times$.
This is illustrated in Figure~\ref{fig:k8}.

We first show in Lemma~\ref{lem:independent} that for every vertex $v^*$, there are words on $\Gamma_0$ whose only unsatisfied vertex is $v^*$.

\begin{lemma}\label{lem:independent}
  For every $v^*\in V_0$ and $e\in\bbF^n$, there is $f\in\W(\Gamma_0,\bbF)$ with $f(v,\cdot)\in\RS[n,k]$ for $v\neq v^*$ and $f(v^*,\cdot)-e\in\RS[n,k]$.
\end{lemma}

\begin{proof}
  We first show that the local views of any $c\in\C_0$ are equal.
  Let $c\in\C_0$.
  For $b\in\bbF_2^r$ and $v\in\bbF_2^r$, denote $\chi_b(v):=(-1)^{b\cdot v}\in\bbF$
  and
  \[
    \hat c_b:=\sum_{v\in V_0}\chi_b(v)c(v,\cdot)\in\RS[n,k]\text.
  \]
  Let $\ell\in[n]$.
  Since $c(v,\ell)=c(v+s_\ell,\ell)$ for all $v\in V_0$, we have
  \[
    \hat c_b(\ell)=\sum_{v\in V_0}\chi_b(v+s_\ell)c(v,\ell)=\chi_b(s_\ell)\hat c_b(\ell).
  \]
  Let $b\neq0$.
  For the $2^{r-1}$ indices $\ell$ such that $b\cdot s_\ell=1$, we have $\chi_b(s_\ell)=-1$, hence $2\hat c_b(\ell)=0$.
  Since the characteristic of $\bbF$ is odd, $2$ is invertible in $\bbF$, so $\hat c_b(\ell)=0$.
  Thus $\hat c_b$ has weight at most $n-2^{r-1}<2^{r-1}$, which is the minimum distance of $\RS[n,k]$, so $\hat c_b=0$.
  Since $2$ is invertible in $\bbF$, so is $2^r$, and the Fourier inversion formula gives for all $v\in V_0$
  \[
    c(v,\cdot)=\frac1{2^r}\sum_{b\in\bbF_2^r}\chi_b(v)\hat c_b=\frac1{2^r}\hat c_0.
  \]
  Therefore all the local views of $c$ are equal to $2^{-r}\hat c_0\in\RS[n,k]$.
  In particular, $c$ is determined by $\hat c_0$, and $\dim\C_0\leq k=2^{r-1}$.

  Now consider the linear map $\Phi\colon\W(\Gamma_0,\bbF)\to\bigl(\bbF^n/\RS[n,k]\bigr)^{V_0}$ such that for $f\in\W(\Gamma_0,\bbF)$,
  \[
    \Phi(f)=\bigl(f(v,\cdot)+\RS[n,k]\bigr)_{v\in V_0},
  \]
  whose kernel is $\C_0$.
  Since $\Gamma_0$ has $2^{r-1}(2^r-1)$ edges, the image of $\Phi$ has dimension at least
  \[
    2^{r-1}(2^r-1)-2^{r-1}=2^r(2^{r-1}-1)=2^r(n-k),
  \]
  which is the dimension of $\bigl(\bbF^n/\RS[n,k]\bigr)^{V_0}$, since $|V_0|=2^r$ and $\dim\bbF^n/\RS[n,k]=n-k$.
  Hence $\Phi$ is surjective.
  In particular, there exists $f\in\W(\Gamma_0,\bbF)$ such that $\Phi(f)$ is equal to $e+\RS[n,k]$ at $v^*$ and to $0$ elsewhere, that is, $f(v^*,\cdot)-e\in\RS[n,k]$ and $f(v,\cdot)\in\RS[n,k]$ for all $v\neq v^*$.
\end{proof}

We then show in Lemma~\ref{lem:far} that such words are far from the code.

\begin{lemma}\label{lem:far}
  If $\U(f)=\{v^*\}$, then $\dH(f,\C_0)\ge\eta_r:=\frac{4^r+4}{2^{r+2}(2^r-1)}>\frac14$.
\end{lemma}

\begin{proof}
  Let $c\in\C_0$.
  Denote $y:=f-c$ and
  \[
    D:=\{v\in V_0\setminus\{v^*\}\mid y(v,\cdot)\neq0\}
  \]
  the set of vertices other than $v^*$ at which the local view of $y$ is nonzero.
  Since $f(v^*,\cdot)\notin\RS[n,k]$ and $c(v^*,\cdot)\in\RS[n,k]$, we have $y(v^*,\cdot)\neq0$.
  Remark that if $y(v,\ell)\neq0$, then $y(v+s_\ell,\ell)=y(v,\ell)\neq0$, so $v+s_\ell\in D\cup\{v^*\}$.
  Applying this to $v=v^*$ and to some $\ell$ such that $y(v^*,\ell)\neq0$, we get $v^*+s_\ell\in D$ since $s_\ell\neq0$, so $D\neq\emptyset$.

  Let $w\in D$.
  Since $w\neq v^*$ and $\U(f)=\{v^*\}$, we have $f(w,\cdot)\in\RS[n,k]$, so $y(w,\cdot)$ is a nonzero codeword of $\RS[n,k]$ and $|y(w,\cdot)|\geq n-k+1=2^{r-1}$.
  As $\Gamma_0$ has no multiple edges, these $2^{r-1}$ nonzero coordinates lead to $2^{r-1}$ distinct vertices of $(D\cup\{v^*\})\setminus\{w\}$, hence $|D|\geq 2^{r-1}$.

  Each edge on which $y$ is nonzero is counted twice in $\sum_{v\in V_0}|y(v,\cdot)|$.
  Therefore $y$ is nonzero on at least half of
  \[
    \sum_{v\in V_0}|y(v,\cdot)|\geq|y(v^*,\cdot)|+\sum_{w\in D}|y(w,\cdot)|\geq1+2^{r-1}|D|\geq1+4^{r-1}
  \]
  edges, out of the $2^{r-1}(2^r-1)$ edges of $\Gamma_0$.
  Thus $\dH(f,c)\geq\eta_r$.
  Finally, $\eta_r>1/4$ since $4^r+4>4^r-2^r$.
\end{proof}

We finally show in Proposition~\ref{prop:main} that a cheating Prover makes the Verifier accept such a word with high probability.

\begin{proposition}\label{prop:main}
  Let $v^*\in V''$, $u^*:=v^*+e_1$, $\ell_0\in L_\times$, and $f$ given by Lemma~\ref{lem:independent} with $e=\one_{\ell_0}$, where $\one_{\ell_0}\in\bbF^n$ is the vector equal to $1$ at index $\ell_0$ and to $0$ elsewhere.
  For every $m\geq 1$, there exists a Prover that makes the Verifier accept $f$ with probability at least $(1-2^{1-r})^m$, whereas, denoting $\delta_r:=\eta_r-2^{1-r}>0$, the bound of Flawed claim~\ref{claim} is smaller than $(1-2^{1-r})^m$ as soon as
  \[
    |\bbF|>\frac{2r^2}{\delta_r\bigl((1-2^{1-r})^m-(1-2^{1-r}-\delta_r/2)^m\bigr)}\text.
  \]
  Hence Flawed claim~\ref{claim} and \cite[Theorem~2]{DMR25} are false.
\end{proposition}

\begin{proof}
  Since $\one_{\ell_0}$ has weight $1<2^{r-1}$, we have $\one_{\ell_0}\notin\RS[n,k]$, hence $\U(f)=\{v^*\}$.

  We first describe the Prover.
  Let $\alpha_0\in\bbF$ be the first challenge of the Verifier, and let $g$ be the word defined for $u\in V'$ and $\ell\in[n]$ by
  \[
    g(u,\ell):=
    \begin{cases}
      \Fold[f,\alpha_0](u,\ell)-\alpha_0 & \text{if }(u,\ell)=(u^*,\ell_0),\\
      \Fold[f,\alpha_0](u,\ell) & \text{otherwise.}
    \end{cases}
  \]
  Since $\ell_0\in L_\times$, $(u^*,\ell_0)$ is a petal of $\Gamma_1$, so $g\in\W(\Gamma_1,\bbF)$.
  Let $u\in V'$.
  If $u\neq u^*$, then neither $u$ nor $\varphi(u)$ is $v^*$, hence
  \[
    g(u,\cdot)=f(u,\cdot)+\alpha_0f(\varphi(u),\cdot)\in\RS[n,k].
  \]
  If $u=u^*$, then by Lemma~\ref{lem:independent},
  \[
    g(u^*,\cdot)=f(u^*,\cdot)+\alpha_0\bigl(f(v^*,\cdot)-\one_{\ell_0}\bigr)\in\RS[n,k].
  \]
  Thus $g\in\C_1$.
  The Prover sends $f_1:=g$, and then $f_i:=\Fold[f_{i-1},\alpha_{i-1}]$ for $i=2,\dots,r$.
  Since each local view of a folding is a linear combination of local views, $f_i\in\C_i$ for all $i\in[r]$.

  We now bound the acceptance probability.
  The checks of rounds $i\geq2$ pass by construction, and the final check passes since $f_r\in\C_r$.
  Since $\Fold[f_0,\alpha_0]$ and $f_1$ differ at most at $(u^*,\ell_0)$, the check of round $1$ fails only if $v_1=u^*$.
  Since $V'$ and $V''$ are disjoint, $v_1=u^*$ if and only if $v_0\in\{u^*,v^*\}$, which happens with probability $2^{1-r}$.
  The $m$ repetitions being independent, the Verifier accepts with probability at least $(1-2^{1-r})^m$.

  Finally, $\delta_r>0$ since $\eta_r>1/4\geq2^{1-r}$ for $r\geq3$ by Lemma~\ref{lem:far}.
  Taking $\varepsilon:=\delta_r/(2r)$ in Flawed claim~\ref{claim} and using $\dH(f,\C_0)\geq\eta_r$, the bound of Flawed claim~\ref{claim} is at most
  \[
    \frac{2r^2}{\delta_r|\bbF|}+\Bigl(1-2^{1-r}-\frac{\delta_r}2\Bigr)^m,
  \]
  which is smaller than $(1-2^{1-r})^m$ under the condition on $|\bbF|$.
  The bound of \cite[Theorem~2]{DMR25} multiplies $\dH(f_0,\C_0)$ by a constant that equals $1$ when $\Gamma_0$ has no petals \cite[Proposition~1]{DMR25}, so it is the same here.
\end{proof}

\end{document}